\documentclass[preprint,proceedings]{rmaa}
\usepackage{paralist}

\usepackage{psfrag,color}

\title{Determination of the mass loss
rate and the terminal velocity of stellar winds. I Genetic algorithm for automatic
line profile fitting}

\author{
  Leonid Georgiev,\altaffilmark{1} 
  Xavier Hernandez,\altaffilmark{1}
}
\altaffiltext{1}{Instituto de Astronom\'\i{}a, UNAM, M\'exico.}

\shortauthor{Georgiev \& Hernandez}
\shorttitle{Genetic algorithm ...}

\fulladdresses{
\item Leonid Georgiev: Instituto de Astronom\'\i{}a, Universidad Nacional
  Aut\'onoma de M\'exico, Apartado Postal 70264, CD Universitaria, CP 04510
  , M\'exico DF, M\'exico DF (\email{georgiev@astroscu.unam.mx}).
\item Xavier Hernandez: Instituto de Astronom\'\i{}a, Universidad Nacional
  Aut\'onoma de M\'exico, Apartado Postal 70264, CD Universitaria, CP 04510
  , M\'exico DF, M\'exico DF (\email{xavier@astroscu.unam.mx}).
}

\listofauthors{Georgiev \& Hernandez}
\indexauthor{Georgiev,L.}
\indexauthor{Hernandez,X.}

\abstract{A new method for automatic fitting of P Cyg line profiles
in UV spectra of stellar winds is presented. The line source function is
calculated using Sobolev approximation and the emergent flux is obtained by
exact integration of the equation of the radiation transport  
(similar to the SEI method described by Lamers et al. (1987)).
The quality of the fit is evaluated using the Likelyhood estimator. The
maximization of the Likelyhood is done by a genetic algorithm. The advantages of
our method with respect to other similar approaches are its robustness and
its insensibility to the initial guess. In addition, the algorithm guarantees the
localization of the global maximum of the Likelyhood hypersurface, which is not 
the case for classical minimization algorithms. Here we present an
implementation of the genetic algorithm for line profile fitting, its tests on both synthetic and real
data and and estimation of the confidence limits of the results.
}

\resumen{
Se presenta un nuevo m\'etodo de ajuste autom\'atico de perfiles P Cyg
observados en espectros UV de vientos estelares. La funci\'on fuente de las l\'{\i}neas
se calcula usando la aproximaci\'on de Sobolev y el flujo emitido 
se obtiene atraves de la soluci\'on exacta de la equaci\'on de transporte (similar
al m\'etodo SEI descrito por Lamer et al. 1987). La calidad del ajuste se
evalua usando el estimador Likelyhood. La maximisaci\'on del Likelyhood
se hace atraves de un algoritmo gen\'etico. Las ventajas de nuestro m\'etodo
comparado con otros m\'etodos similares son su estabilidad y su poca sencibilidad de
las condiciones iniciales. Ademas, el m\'etodo garantiza la localizaci\'on del maximo global de la superficie
del Likelyhood, cual no se puede hacer con los algoritmos cl\'asicos.
Presentamos la implementaci\'on del metodo, las pruebas de su funcionalidad
usando datos sint\'eticos y reales y la estimaci\'on de los rangos de confianza
de los resultados.
}

\addkeyword{}
\addkeyword{}
\addkeyword{}
\addkeyword{}

\begin{document}
% Typeset article header
\maketitle

\section{Introduction}
	Many hot stars lose mass through a supersonic wind. The terminal velocity of these winds, 
and the details of the mass loss rate  play an important role in stellar evolution. Stellar winds 
crucially determine the interaction of the star with its surrounding interstellar medium, this being 
one of the central feedback mechanisms, which in turn modulate star formation at galactic levels. 
Estimating stellar mass loss rates is therefore a highly relevant problem.
	
The best way of determining the mass loss rate of a star is by studying the radio emission 
from the wind itself (Lamers et al. 1999) however, with current observational techniques this is 
limited to nearby stars with high mass loss rates. The other reliable method is based 
on a detailed modeling of the stellar spectrum. The recent development of realistic codes 
which include many atomic lines makes the latter approach the preferred one. 
Unfortunately, the fitting of the spectrum requires high quality data which covers a wide range of 
wavelengths and with high dispersion, this again limits the method to bright or  nearby stars. 
An alternative has been proposed by Lamers et al. (1987) and Groenewegen \& Lamers (1999). 
This method is based on a code which approximates the opacity of the wind as a 
simple function of the distance from the stellar nucleus. The code calculates the source 
function in a given line within the Sobolev approximation (Sobolev, 1957) and then 
calculates accurately the emitted profile. The fit of the observed profile gives the terminal 
velocity of the wind together with the total optical depth along the
line of sight.  The mass loss rate is estimated from the later (see next section). 
Even though the obtained value is sensitive to certain assumptions regarding  
the ionization fraction of the element whose line is fitted, the method has the advantage of being applicable 
to faint stars for which only low resolution and low signal to noise spectra are available.

The main disadvantage of this approach is the large number of free parameters which have to be adjusted. 
Groenewegen \& Lamers (1989) proposed a method for determination of the parameters 
by a "classical" minimization of the $\chi^2 = \sum\left(1-F_o/F_c\right) $ where $F_o$ is the 
observed flux and $F_c$ is the calculated one. But the multi-dimensional fitting converges to the 
correct solution only if the initial guess is close to the $\chi^2$ minimum. The complex 
topology associated with the high dimensional hypersurface being treated leads to the appearance of
numerous local minima. The ``classical'' minimization methods usually get trapped in the local minimum close
to the initial guess missing the global one. The somewhat lengthy numerical procedure involved, again 
together with the large number of dimension of the problem, makes a dense sampling of the parameter space 
completely impractical. Recent years have seen the development of 
non-standard maximization techniques such as simulated annealing and genetic algorithms, particularly in
connection with astrophysical applications (e.g. Teriaca et. al 1999 in fitting radiative transfer 
models to solar atmosphere data; Sevenster et al. 1999 in fitting parameters of Galactic structure 
models to observations or Metcalfe 2003; for the fitting of White dwarf astroseismology parmeters).
These methods provide a framework which in a reliable
manner identifies the global maxima (or minima, as the case might be) of complex multi-dimensional 
surfaces, with a close to optimal use of computing resources.

In this paper we present an application 
of a genetic algorithm used to obtain the optimal parameters of a stellar wind, within the Sobolev 
approximation, for the fitting of P Cyg line profiles of UV resonance lines. 

Section (2) describes the radiative transfer model. In section (3) a description of the genetic algorithm
is given, with use of it in a number of synthetic test cases and real data examples appearing in section (4).
Section (5) presents our conclusions.

\section{The algorithm}

\subsection{Radiative transfer code} 
	The core of the method is a code which solves the radiative transport problem under the
assumption of a spherically symmetric stellar wind. Following the SEI code (Lamers et al. 1987) we 
approximate the radial Sobolev optical depth $\tau(v)$ as:

\begin{equation}
\tau(v) = \frac{\chi_0 c}{\nu_0 \frac{dv}{dr}} = T \, v^{\alpha_1} \, (1-v)^{\alpha_2}
\label{eq_1}
\end{equation}

where $v=\frac{V(r)}{V_\infty}$ is the velocity of the wind, normalized to the terminal velocity $V_\infty$, $\chi_0$
is the central opacity of the line, $\nu_0$ is the laboratory frequency of the transition and
$c$ is the speed of light. $T$, $\alpha_1$ and $\alpha_2 $ are free parameters of the model,
which describe the  radial dependence of the optical depth. The velocity field of 
the wind is adopted to be a standard $\beta$-law

\begin{equation}
V(r) = V_\infty \, \left(1-\frac{R_0}{r}\right)^\beta
\label{eq_2}
\end{equation}

where $V_\infty$ is the terminal velocity. The winds of the hot stars are not smooth
but have random motions. The shape of the
blue wing of the P Cyg absorption component suggest that the intrinsic line profile
has a width on the order of hundreds of kilometers per second.  The physical nature of these
random motions is not clear but they act as an additional line broadening mechanism similar to 
turbulence and are frequently called "turbulence" although they may  have little to do with it.
We model this chaotic motions assuming the intrinsic line profile to be Gaussian defined as

\begin{equation}
\phi(v) = \frac{1}{\sqrt{pi}V_{turb}} exp^{-\left(\frac{v}{V_{turb}}\right)^2}
\end{equation}

The turbulent velocity, $V_{turb}$, can be variable throughout the wind, but because its physical nature
is not clear there is no model of its possible variability. To keep the model as
simple as possible, we assume $V_{turb}$ to be constant and a parameter of the model.

Finally, we set the innermost point of the grid, $R_{min}$ to the location where $V(R_{min})$ is equal to the
sound speed $V_{sound}$. This is done by setting the parameter $R_0$ in (\ref{eq_2}) as
\begin{equation}
R_0 = R_{min} \left[1.0-\left(\frac{V_{sound}}{V_\infty}\right)^{\frac{1}{\beta}}\right]
\label{eq_3}
\end{equation}

Equations (\ref{eq_1}), (\ref{eq_2}) and (\ref{eq_3}) determine the distribution of 
the central opacity of the line $\chi_0$ as a function of the radius $r$. 
Within the Sobolev approximation, one can now calculate the line source function $S$ and 
then can apply a formal solution of the equation of radiative transport throughout the 
wind and calculate the emergent flux (see Georgiev \& Koenigsberger, 2004 for more details 
of the code). The code is designed to work in 3D geometry, but for the current 
test purposes, the solution is restricted to the case of spherical symmetry. Finally, we calculate the
mass loss rate as follows (Groenewegen \& Lamers, 1989). The line opacity $\chi_0$ is
\begin{equation}
\chi_0 = \frac{\pi e^2}{mc} f_{lu} n_l = \tau(v)*\frac{\nu_0}{c} \frac{dV(r)}{dr},
\label{eq_chi}
\end{equation}
where $n_l$ is the population of the lower level, $f_{lu}$ is the oscillator strength and
we do not take into account the correction for stimulated emission. The level population $n_l$
can be written as
\begin{equation}
n_l = N_{ion} q_{ex} = N_{atom} q_{ion} q_{ex} = N_H A_E q_{ion} q_{ex},
\label{eq_nl}
\end{equation}
where $q_{ion}$ and $q_{ex}$ are the ionization and excitation fractions and $A_E$
is the chemical composition of the element relative to Hydrogen by number.  Then the level population $n_l$
is related to the mass loss rate by the equation of continuity
\begin{equation}
n_l = \frac{A_E}{\mu} \frac{\dot M q_{ion} q_{ex}}{4 \pi r^2 V(r)},
\label{eq_nl1}
\end{equation}
where $\mu$ is the average molecular weight. Substituting (\ref{eq_nl1}) into (\ref{eq_chi}), one can
obtain the product $\dot M q_{ion} q_{ex}$ at each point $r$ as
\begin{equation}
\dot M q_{ion} q_{ex} = \tau(v) \frac{dV(r)}{dr} r^2 V(r) \frac{\mu}{A_E} \frac{4 m c}{e^2 f_{lu}} \frac{\nu_0}{c}.
\label{eq_mdot}
\end{equation}

Then, if the parameters $V_\infty,V_{turb},\beta,T,\alpha_1$ and $\alpha_2$ are given, equation (\ref{eq_1})
and (\ref{eq_2}) together with the atomic data and the chemical composition allow the estimation of the
product $\dot M q_{ion} q_{ex}$. The method described in this paper suggests a better way to obtain the six 
parameters that describe the velocity law and the optical depth. It may not improve the precision of the calculated
mass loss rate if the dominant uncertainty is the ionization fraction.

Applying the above procedure, we are able to calculate any line profile, once the six parameters of the method
$V_\infty,V_{turb},\beta,T,\alpha_1$ and $\alpha_2$ have been specified. The next step is the selection of a
statistical estimator of the goodness of fit to be associated to a particular set of parameters and a given
observed profile. From a Bayesian perspective, the optimal such indicator is the Likelyhood,
calculated as the probability that a certain data set be observed, given a model. We hence compare a modeled 
profile against a given observation through a Likelihood estimator defined as: 

\begin{eqnarray}
\log L(V_{\infty}, V_{turb}, \beta, T, \alpha_{1}, \alpha_{2}) = && \label{eq_4}\\ \nonumber
\sum_{i} \log \left[ {1 \over {(2 \pi)^{1/2}  \sigma_{i}}} exp-\left(\frac{F_{i,o} - F_{i,c}}
{2 \sigma_{i}}\right)^2 \right]
\end{eqnarray}. 

Where $F_{i,o}$ and $F_{i,c}$ are the observed and calculated flux respectively at a certain wavelength, 
$\lambda_{i}$ and the factors $\sigma_{i}$ are the observational errors at the same $\lambda_i$. 
The weight $\sigma$ assigned to each observed point is taken from the signal-to-noise ratio 
of the observations. Setting a large $\sigma_i$ to certain points forces them to participate less
in the final fit, which allows us to exclude effectively spectral regions where overlap with 
other lines occur. Hence, the final result depends on the assigned $\sigma_i$. In practice, the bands of the
observed spectrum affected by spurious details are easily detectable and the assignment of the
weights is straightforward.

The synthetic profile is calculated over a grid adjusted for better sampling the 
velocity field in the wind. The calculated profile is then interpolated on the wavelengths
of the observed profile using monotonic cubic polynomials (Steffen, 1990). Finally, equation
(\ref{eq_4}) is calculated and used as a measure of the quality of the fit.

\subsection{The genetic algorithm}

Once we have a method for calculating the line profile which corresponds to a particular set of six
parameters, and an estimator of the goodness of the fit, we must now determine what the optimal 
parameter vector is; i.e. we must find the combination of six parameters which maximizes our Likelihood 
function. This will yield the physical parameters of the stellar wind associated to the observed line profile, 
from which the mass loss rates and wind terminal velocities can be estimated.

A dense exploration of our six-dimensional parameter space 
is unfeasible. Classical maximization techniques are also unreliable in connection to problems of such 
high dimensionality, and liable to get trapped by local maxima (Press et al . 1989). In fact we have 
encountered numerous test cases where local maxima exist in the Likelyhood surface of our problem. 
We bypass these obstacles using genetic algorithms in searching for the maximum
of the Likelyhood surface.

The idea is to simulate a population of organisms (we call them "bugs") which breed and evolve following
prescriptions based on that
biological systems are thought to follow, the result being a progressive increase in the fitness of the
population, with the fittest individual in each generation eventually reaching the absolute maximum of the
fitness surface. For the case at hand, a "bug" is a set of wind parameters described above.

Once an observed line profile has been picked, the first step is to select $N$ random points in our 
parameter space, each a parameter vector
for the stellar wind, ($V_{\infty}, V_{turb}, \beta, T, \alpha_{1}, \alpha_{2}$). The ranges over which
these parameters are to be chosen have to be determined by inspection of the observed line profile. Only ample 
margins enclosing the expected value are needed. We turn to this issue in the examples section.

The six coordinates of each point are then turned into binary numbers, with a pre-determined resolution
and then combined in a string. We have used 8 bits per parameter for our implementation, resulting 
in a string of $N_G = 6$ parameters $\times 8$ bits $= 48$ bits. This selection gives to our search algorithm a resolution of 
$Range_{i}/2^{8}$ in each of the six dimensions where $Range_{1-6}$ are the ranges for each parameter
chosen above.

Each of the $N$ randomly selected points becomes a ``bug'' of the first generation. The string
of $N_G$ $1^{s}$ and $0^{s}$ which corresponds to its 6 coordinates in parameter space becomes its ``genotype'',
and the Likelyhood associated to that point becomes its ``phenotype''. The $N$ members of the first generation 
are then ranked according to the value of the Likelihood associated to each, with those ranking above
$N/S_P$ being deemed ``superior'', and those ranking below this threshold ``inferior''. $S_P$ is a parameter
of the simulation which has to be a number greater than 1.0, typically around 1.5.

This completes the first generation, which are now ``mated'' to produce the second one. This proceeds
by the random selection of pairs of individuals from the pool of $N$ members of the first generation.
Once a pair is selected, ``offspring'' are calculated. The first $C_P$ genes of one of the
parents are taken for first $C_P$ genes of the offspring, and the remainder are taken from the other parent. 
$C_P$ is a ``crossing point'', chosen at random in the interval (0-$N_G$), each time an offspring is constructed.

In this way, the offspring will have ``genes'' in some aspects resembling those of one parent, and in some
cases the other, the corresponding position in parameter space will hence share some coordinates with one
parent, and some with the other, with one of them being a mixture. Also a fraction $M_F$ of 
randomly selected genes in the
new generation are ``mutated'' i.e. they are flipped from 1 to 0, or 0 to 1, as the case might be.
$M_F$ is a second and last parameter of the method, which together with $S_P$ must be chosen and tuned
through some testing of the algorithm in the context of the particular problem being treated. We have used
$M_F$ close to 0.01 which means 1\% of the genes are mutated

The number of offspring to be produced by each pair depends on the fitness of the parents, if both are
``superior'' individuals, 3 offspring are calculated. The pairing of a ``superior'' and an ``inferior'' 
individual yields only 2 offspring, and when two unfortunate ``inferiors'' mate, only one child ensues.
The process is repeated until N offspring have been obtained, the new generation has been constructed,  
and it completely replaces the former. 

The parameter $S_P$ hence represents a selection pressure. If chosen close to 1, most individuals will 
be deemed ``superior'', and hence the second generation will be distributed in parameter space
much like the first one was. However, as $S_P$ takes larger values, the second generation will be made up 
of the sons of (mostly) the ``superior'' individuals of the former, yielding a gradual climb up the Likelihood 
surface of the problem. 
Next the genotypes of the new generation are turned back into coordinates in our parameter space, and then 
into phenotypes -the associated Likelihood. The ranking is repeated, and the cycle iterated.

In this way, we have introduced mating, selection, and chance mutations, the three main ingredients of
biological evolution, albeit in a highly simplified manner. Selection forces subsequent generations
up the Likelihood surface, with the random mutations ensuring that a fraction of the individuals are constantly
sampling new regions, which in turn guarantees the absolute maximum will eventually be found.

In many classical maximization algorithms one is called to evaluate the gradient of the surface at any given
point. When the surface is not an analytical one, but the result of a lengthy
numerical procedure, as in our case, the gradients could quite easily be dominated
by roundoff errors. The method described calls for no gradients of the surface, only its value at the points
being tested. With the resolution described here, a dense sampling of our Likelihood surface would have
required $(2^{8})^{6}=2.8  \times 10 ^{14}$ evaluations of our wind model. We have used N=100 members per
generation, and both in synthetic examples where the answer was known in advance, and in tests with real data,
have found convergence in 200 generations, i.e. only $2 \times 10^{4}$ costly line profile calculations.

The method described here differs only slightly from the general purpose genetic maximization algorithm
PIKAIA, described in Charbonneau (1995), and having been used successfully in a number of astrophysical 
applications to date (e.g. Teriaca et. al 1999, Metcalfe 2003). The differences between the two are limited to the way 
``selection'' is treated, the implementation described here giving results more suitable for the particular
problem we are treating.

\section{Test cases}

\begin{figure*}[!t]
  \includegraphics[width=16cm,height=8cm]{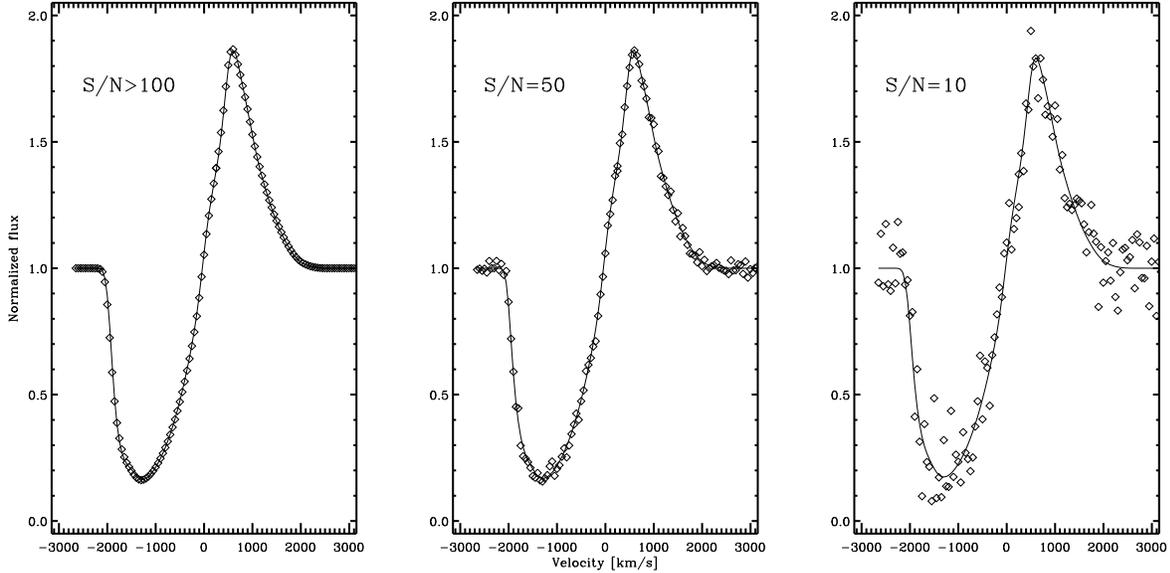}%
  \caption{Test of the fitting algorithm. A synthetic profile of C IV 1548/1550\AA \,
  (squares) is shown together with the best fit (solid line) for three adopted S/N ratios.}
  \label{fig_synt}
\end{figure*}

	The first test of the algorithm is made on 
synthetic data. We calculate the line profile of the C~IV~1548/1550\AA \, doublet
with a predefined velocity law and wind density distribution and then fit this
profile using the genetic algorithm. Fig.~\ref{fig_synt} shows the original profile and the
resulting fit for three different signal-to-noise ratios. 
The parameters used to generate the profile and the result of the
fitting are summarized in Tab.~\ref{tab_synt}.

\begin{figure}[!t]
  \includegraphics[width=\columnwidth]{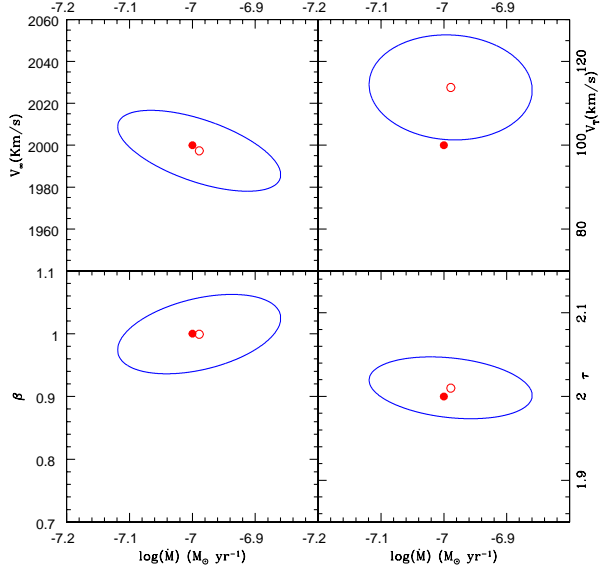}%
  \caption{The ellipses show $1-\sigma$ confidence intervals on the recovered parameters (circles)
 for the fit to the synthetic profile with S/N=100. The input values are shown by the solid dots.}
  \label{fig_el100}
\end{figure}

\begin{figure}[!t]
  \includegraphics[width=\columnwidth]{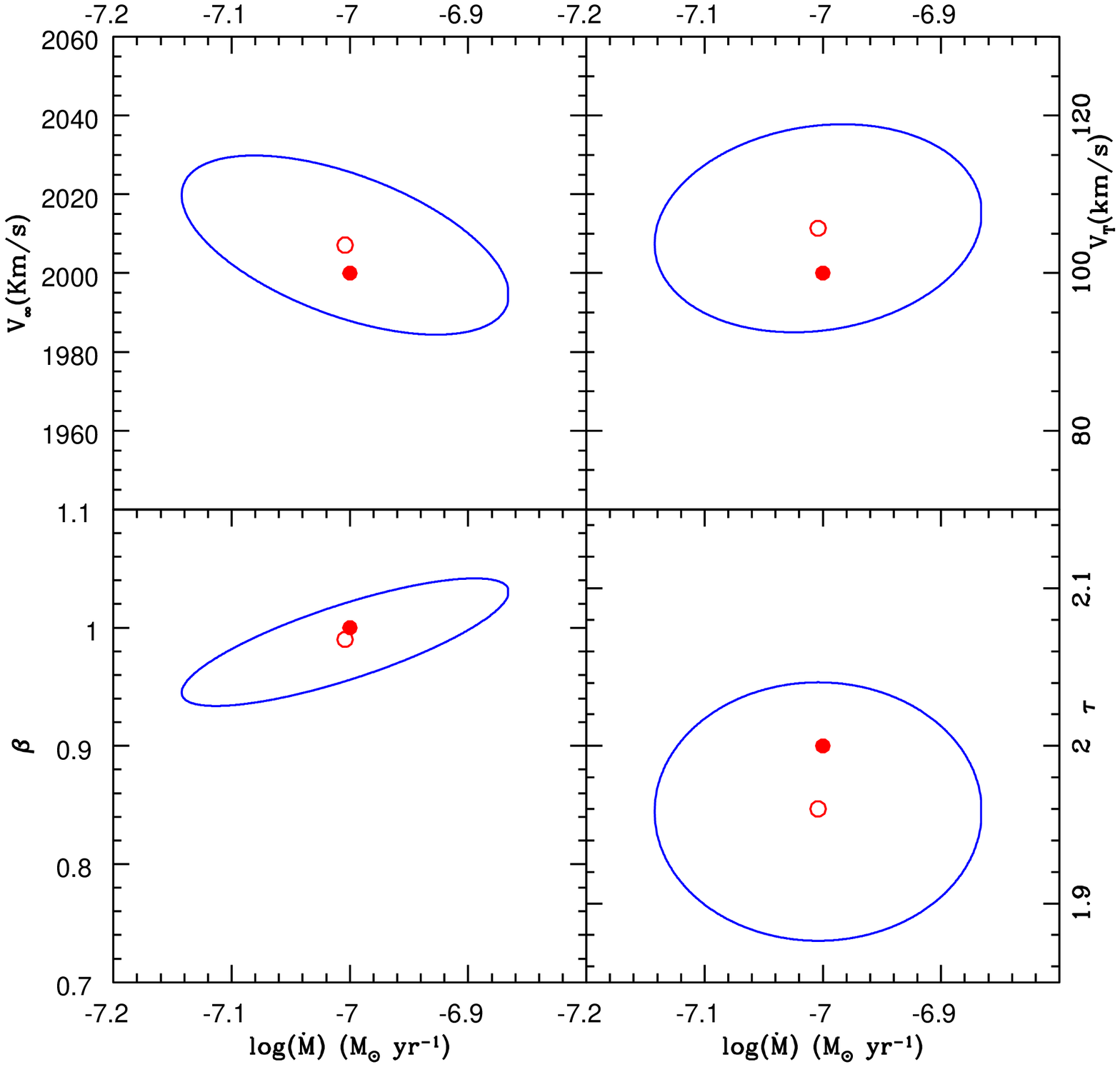}%
  \caption{The ellipses show $1-\sigma$ confidence intervals on the recovered parameters (circles)
 for the fit to the synthetic profile with S/N=50. The input values are shown by the solid dusts.}
  \label{fig_el50}
\end{figure}

\begin{figure}[!t]
  \includegraphics[width=\columnwidth]{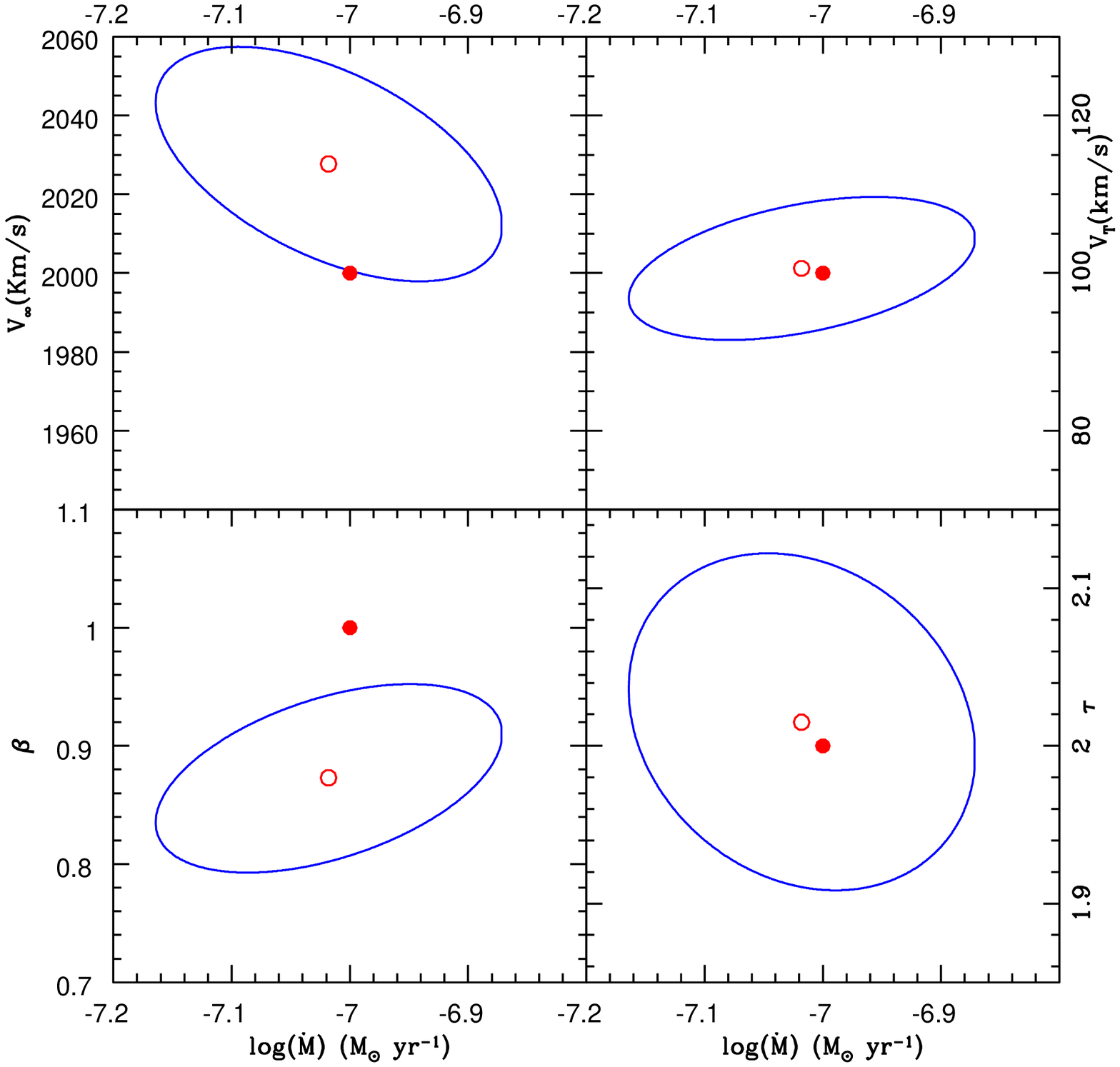}%
  \caption{The ellipses show $1-\sigma$ confidence intervals on the recovered parameters (circles)
 for the fit to the synthetic profile with S/N=10. The input values are shown by the solid dusts.}
  \label{fig_el10}
\end{figure}

	The genetic algorithms do not include  an easy way of
estimating the errors of the results. The method only serves to locate the global maximum in
an efficient manner. We estimate the confidence intervals of the recovered parameters by Monte-Carlo simulations
of synthetic line profiles. In this way, all uncertainties inherent to the method and the modelled data are taken
into account. We select three signal-to-noise ratios: 100, 50 and 10. For each of them we generate 20 different line profiles, using the same synthetic data but adding different and independent noise. The obtained line profiles
were fitted by the genetic algorithm using different first generations of "bugs". Figs.~\ref{fig_el100},\ref{fig_el50} and \ref{fig_el10} show the
confidence intervals obtained through this method for the recovered parameters for the simulated
profiles of Fig.~\ref{fig_synt}


Once the Monte-Carlo simulations are all done, we find the means for all the recovered parameters, and of the
recovered mass loss rates. Next, a full covariance analysis is performed, yielding the parameter's standard 
deviations, which together with the relevant covariances can be used to construct
statistically meaningful confidence ellipses. Fig.~\ref{fig_el100} shows the $1-\sigma$ ellipses for the 
case of S/N =100, with the solid dots indicating the input values, and the circles showing the ones reached by the fitting algorithm.
The mass loss rate is recovered almost exactly, with a $1-\sigma$ interval of 30\%. A negative correlation is
seen between the inferred $\dot M$ and the recovered terminal wind velocity, while mass loss rate
and $\beta$ show a weak positive correlation. Figs.~\ref{fig_el50} and \ref{fig_el10} are equivalent to
 Fig.~\ref{fig_el100}, but for the cases with S/N values of 50 and 10, respectively. The same qualitative
features can be seen, with the ellipses getting only slightly larger, to reach 40\% for the S/N =10 case.
It is reassuring that errors on the inferred mass loss rate are not highly sensitive to the value of the S/N
of the simulated profile, making the method usefull even in cases where observations are not of the
best quality. Also, we note that no systematics appear on the recovered $\dot M$, we have constructed
unbiased estimator of the important physical parameter we set out to infer. We have to stress that the 
error on the mass loss rate are the formal errors of the fit. The errors related to the determination
of the ionization fraction will be treated in the next paper of the series.

\begin{table*}[!t]\centering
  \setlength{\tabnotewidth}{\columnwidth}
  \tablecols{5}
  % Stretch the space between table columns 
  \setlength{\tabcolsep}{2.8\tabcolsep}
  \caption{ Parameters of the synthetic profile and the results of the fits.}
  \begin{tabular}{lrrrr}
  \toprule
    Parameter & Exact value & S/N=100 & S/N=50 & S/N=10 \\
  \midrule
    $V_\infty$ (km s$^{-1}$)  & 2000 & 1997 $\pm$ 19 & 2007 $\pm$ 23 & 2014 $\pm$ 30\\
    $V_{turb}$ (km s$^{-1}$)  &  100 &  114 $\pm$ 12 &  107 $\pm$ 13m&  101 $\pm$ 9 \\
    $\beta$                 &  1.0 & 0.99 $\pm$ 0.06&  0.99 $\pm$ 0.05& 0.87 $\pm$ 0.08\\
    T                       &  2.0 & 2.01 $\pm$ 0.06 & 1.96 $\pm$ 0.08& 2.015 $\pm$ 0.11\\
    $\log \dot M q_{ion} q_{ex}$ (M$_\odot/yr$)           & -7.0 & -7.00 $\pm$ 0.13 & -7.00 $\pm$ 0.14 & -7.02 $\pm$ 0.15\\
  \bottomrule
  \end{tabular}
  \label{tab_synt}
\end{table*}

\begin{figure}[!t]
  \includegraphics[width=\columnwidth]{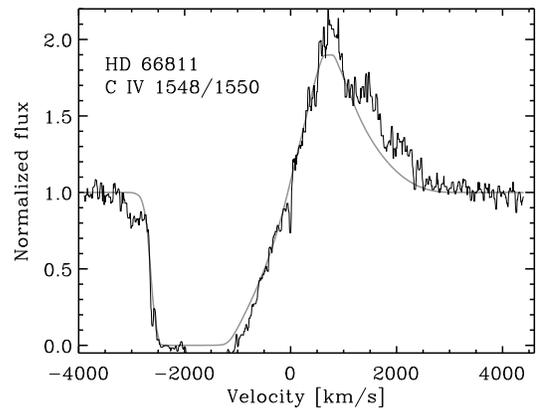}%
  \caption{Fit of the C IV 1548/1550\AA\, line in $\zeta$ Pup (=HD66811).
  The thin line is the observed spectrum. The dotted  line is
  the fit}
  \label{fig_civ}
\end{figure}

\begin{figure}[!t]
  \includegraphics[width=\columnwidth]{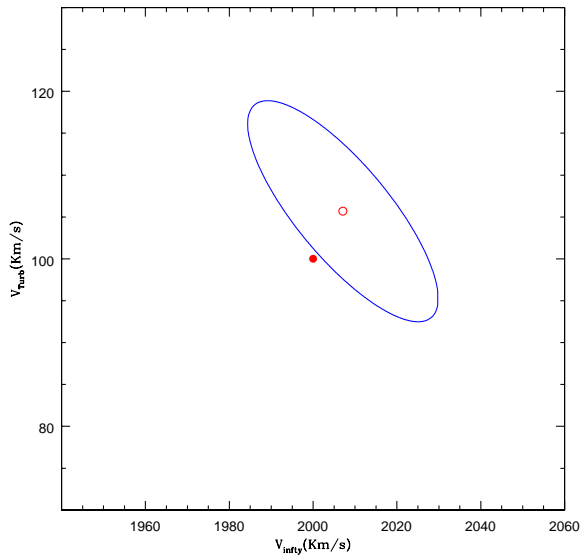}%
  \caption{Correlation between $V_\infty$ and $V_{turb}$ calculated for 
  synthetic profile with S/N=50.}
  \label{fig_corel}
\end{figure}

\begin{figure}[!t]
  \includegraphics[width=\columnwidth]{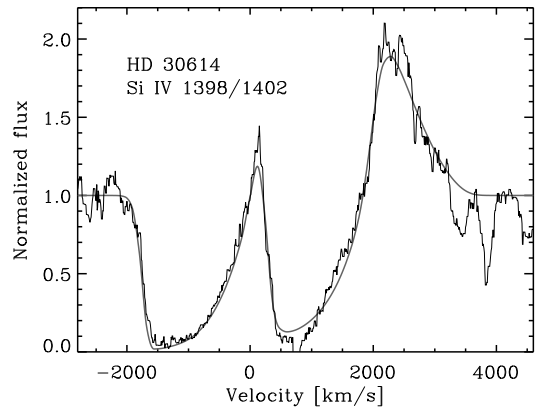}%
  \caption{Fit of the Si IV 1392/1402\AA\, line in HD~30614. The dotted line is the fit}
  \label{fig_siiv}
\end{figure}

	As can be seen from Tab.~\ref{tab_synt} in all cases the algorithm finds the correct maximum
of the likelihood estimator, the mass loss rate is recovered successfully, always well within $1-\sigma$ of
the input value, even at low S/N values. As an additional check we tried to fit the line profiles
of real stars. Fig.~\ref{fig_civ} shows the profile of C IV 1548/1550\AA\, doublet observed in IUE spectrum
of $\zeta$ Pup (HD668110) and Fig.~\ref{fig_siiv} shows the profile of Si IV 1398/1402\AA\, doublet in the
IUE spectrum of HD~30614.
The comparison between the results of our fit and the parameters obtained 
by Groenewegen \& Lamers (1989) (Tab.~\ref{tab_obs}) shows a good agreement. Our solution shows a slower 
velocity law (large $\beta$) but our final fit has higher Likelyhood than the fit obtained with the published parametes. For both stars we obtain a
larger $V_\infty$ and lower $V_{turb}$. As shown in Figs.~\ref{fig_corel}, the
two parameters are correlated and one could expect this behavior. But in both cases the automatic fit gives 
values very close to the more human controlled solution.

\begin{table*}[!t]\centering
  \setlength{\tabnotewidth}{\columnwidth}
  \tablecols{5}
  % Stretch the space between table columns 
  \setlength{\tabcolsep}{2.8\tabcolsep}
  \caption{ Parameters of the wind of HD30814 and HD66811 determined in this work
  together with the same data obtained by Groenewegen \& Lamers (1989).}
  \begin{tabular}{lrrrr}
  \toprule
    Parameter & \multicolumn{2}{c}{HD66811 C IV 1548/1550\AA} & \multicolumn{2}{c}{HD30814 Si IV 1398/1402\AA} \\
              & This work & Published & This work &  Published \\
  \midrule
    $V_\infty$ (km s$^{-1}$)  &  1550 $\pm$ 16   & 1550 $\pm$ 50  & 2137  $\pm$ 21   & 2200 $\pm$ 60 \\
    $V_{turb}$ (km s$^{-1}$)  &   240 $\pm$ 24   &  190 $\pm$ 70  &  287  $\pm$ 30   & 290 $\pm$ 70  \\
    $\beta$                 &  1.13 $\pm$ 0.1  &  0.7 $\pm$ 0.1 & 1.18  $\pm$ 0.1  & 0.7 $\pm$ 0.1 \\
    $\log \dot M q_{ion} q_{ex}$ (M$_\odot/yr$)          & -6.99 $\pm$ 0.15 &  $>$ -7.14     & -7.27 $\pm$ 0.15 & $>$ -8.4      \\
  \bottomrule
  \end{tabular}
\label{tab_obs}
\end{table*}

\section{Conclusions and Discussion} 
        The determination of the basic stellar parameters like wind velocity and 
mass loss rate is a very time consuming process. In this paper we present a method 
which allows an objective determination of these parameters based on a completely
automatic procedure. The test cases shown above support the robustness of the method
even in case of low S/N data. The method is especially useful for objects for which only low
resolution and low S/N data are available. Their spectra show only few 
spectral features and the application of full NLTE codes is not practical. Our method was
successfully applied by Arrieta \& Stangellini (2004) for central stars of LMC planetary
nebulae. Also, the use of Monte-Carlo simulations on the recovered parameters allows for the secure
and objective determination of statistically meaningfull confidence intervals around recovered parameters,
indeed, the full covariance matrix is available. But, we have to stress, that the parameter obtained 
by this method is not the mass loss rate $\dot M$ itself, but rather the product
$\dot M q_{ion} q_{ex}$. 
There is no easy way to separate these quantities on the base of low dispersion and low S/N data
on which only few lines are measurable. Our method helps to obtain a more reliable estimates of the product
in an easier way, but the
final decision on how to disentangle $\dot M$ and the ionization and excitation fractions depends on the 
studied object and on the problem which needs to be solved. 

An important advantage of our method, compared to other fitting 
algorithms is a minimal effect of the initial guess on the final solution. This is because the method 
does not rely on a series expansion. The genetic algorithm requires only a crude margins on the parameters 
to find the maximum of the likelyhood surface. In case the maximum is not within the proposed
ranges, the population migrates to the parameter's limit which is closer to the maximum. A subsequent run
with new ranges that increase this limit usually leads to a successfully solution.
The price to pay is only an increased number of profile evaluations. But the fast computers make this price 
affordable. 
The calculation of one generation of "bugs" takes less than 2 minutes on a Pentium IV class computer. Usually 
fifty to a hundred generations are needed to fit one line profile. Each "bug" in a generation is independent 
of the rest of the population, so a simple parallelization increases the speed of the calculation almost 
proportional to the number of available processors. 

\begin{figure}[!t]
  \includegraphics[angle=90, scale=0.5]{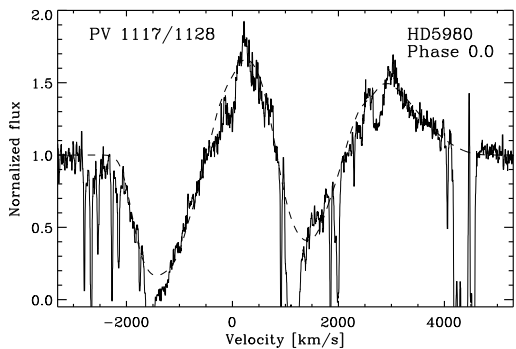}%
  \caption{Fit of the P V 1117/1129 line in the LMC star HD5980.
  The thin line is the observed spectrum. The dashed line is 
  the fit. Small weight were applied to the interstellar absorption
  line, which contaminate the P Cyg profile.}
  \label{fig_pv}
\end{figure}

        The method has however few disadvantages. First of all, the choice of the weights $\sigma_i$ in
equation (\ref{eq_4}) is not automatic. The observed profiles frequently have overlapped absorption lines
of other elements, interstellar lines or absorption components of the same star and the same element
but which are not formed in the wind. The weights $\sigma_i$ should be chosen lower in the
affected bands so the Likelyhood is calculated using only the pixels which belongs to the line.
Fig~\ref{fig_pv} show the fit of the P V 1117/1128\AA\, line in the star HD5980 as an example of
the performance of the code on a highly contaminated line. The FUSE spectra are characterized by
a huge number of interstellar absoption line. In the above case, a small weight (large errors) were
assigned to the bands with high contamination. As seen from Fig~{\ref{fig_pv}, the code finds 
a satisfactory fit which reproduce the P Cyg line profile, avoiding the contamination.

Another problem is that the parameters can be obtained only from a grid. 
As a result, the code never converges to the exact solution but to the
node of the parameter's grid which is closest, but not exactly at, the maximum of the likelyhood 
surface. That might cause the systematic
error seen in Fig.~\ref{fig_el10}. One possible solution to this problem is
a second run of the code with smaller ranges of the parameters around the solution found. 
Another approach is to combine a genetic algorithm with a classical minimization method. The genetic
algorithm guarantees, that the solution is near the global maximum of the likelyhood surface and then
the classical minimization algorithm starts from that point and finds the exact solution. Until now we 
have explored only the first option. The combination of the algorithms and their application
to a wide range of objects, together with the problem of
separating the mass loss rate from the ionization fraction will be subject of the next paper of this series.  
 
\acknowledgments
We thank Gloria Koenigsberger for many usefull discutions and for critical reading of the manuscript.
This work was supported by CONACyT grants 40864 and 42809 and UNAM/DGAPA grant IN107202.

\end{document}